\documentclass[11pt]{article}
\usepackage{latexsym}
\usepackage[dvips]{graphics,color}
\usepackage{graphpap}
\usepackage{alltt}
\usepackage{graphicx}
\usepackage{amsfonts}
\usepackage{amsmath}
\usepackage{amsbsy}
\usepackage[active]{srcltx}
\usepackage{psfrag}
\usepackage[pdftex, colorlinks=true, linkcolor=blue, citecolor=red]{hyperref}
\usepackage{xcolor}

\newtheorem{theorem}{Theorem}[section]
\newtheorem{lemma}[theorem]{Lemma}
\newtheorem{proposition}[theorem]{Proposition}

\newtheorem{definition}[theorem]{Definition}

\newcommand{\beq}{\begin{equation}}
\newcommand{\feq}[1]{\label{#1} \end{equation}}
\newcommand{\beqr}{\begin{eqnarray}}
\newcommand{\feqr}{\end{eqnarray}}
\def\non{\nonumber}

\newcommand{\rf}[1]{(\ref{#1})}

\DeclareFontFamily{U}{eufm}{}
\DeclareFontShape{U}{eufm}{m}{n}{<->eufm10}{}
\DeclareSymbolFont{mcy}{U}{eufm}{m}{n}
\DeclareMathSymbol{\Hr}{\mathord}{mcy}{"58}

\def\mas#1#2#3{Meth. Appl. Sci. {\bf{#1},} #2 (#3)}

\def\pams#1#2#3{Proc. Amer. Math. Soc. {\bf{#1},} #2 (#3)}

\def\dok#1#2#3{Dokl. AN SSSR {\bf{#1},} #2 (#3)}
\def\sam#1#2#3{Stud. Appl. Math. {\bf{#1},} #2 (#3)}
\def\cpam#1#2#3{Commun. Pur. Appl. Math {\bf{#1},} #2 (#3)}
\def\arma#1#2#3{Ark. Mat. {\bf{#1},} #2 (#3)}
\def\dans#1#2#3{Dokl. Akad. Nauk SSSR {\bf{#1},} #2 (#3)}

\begin{document}


\begin{center}


{\Large \bf The Dirichlet kernel on the real hyperbolic space for radial functions}\\
[4mm]

\large{Agapitos N. Hatzinikitas} \\ [5mm]

{\small Department of Mathematics, \\ 
University of Aegean, \\
School of Sciences, \\
Karlovasi 83200\\
Samos, Greece \\
E-mail: ahatz@aegean.gr}\\ [5mm]

\end{center}

\begin{abstract}
Asymptotic expansions as well as necessary and sufficient conditions are provided for the pointwise convergence of the spherical partial integrals of the associated Fourier transforms on the real hyperbolic space. The proposed method permits one to produce new results and opens a connection with already known ones utilizing a unified framework.  
\end{abstract}

\noindent\textit{PACS 2010:02.30.N, 02.30.G, 02.50.S, 02.30.M}   
\section{Introduction}
\label{sec0}
Pinsky in Refs. \cite{PIN1,PIN2} first established necessary and sufficient conditions for piecewise smooth functions, supported on a ball, which guarantee the convergence of the spherical partial integrals at a pre-assigned point on the $\mathbb{R}^d, \, \mathbb{T}^d$ and $\mathbb{H}^d$ spaces. 
\par In the present work, for spherically symmetric functions we adopt the geodetic spherical polar realization of $\mathbb{H}^d$ and allow the pole of the hyperboloid to move freely along the positive $X^0$-axis, thus opening a connection with the Euclidean geometry at infinity. A consequence of this formulation is that it enables us to treat the problem on an equal footing  for both $\mathbb{H}^d$ and $\mathbb{E}^d$ spaces. The paper more precisely is organized as follows.
\par In Sec. \rf{sec1} we give a brief overview of geometrical aspects related to the real hyperbolic space as well as some properties of the hyperbolic plane waves $\mathcal{E}_d(x,p)$. In the sequel, the spherical mean-value $\Phi_{\lambda}(\chi)$ of $\mathcal{E}_d(x,p)$ is reduced to an integral in one variable, using the Funk-Hecke integral identity. Finally, $\Phi_{\lambda}(\chi)$ is expressed in terms of the Legendre and Gauss hypergeometric functions. This identification to known functions will be very useful in proving asymptotics in the next section. We conclude this section with a known theorem concerning the Fourier-Helgason transform for radial functions on the real hyperbolic space.
\par In Sec. \rf{sec4} we define the Dirichlet kernel and in Proposition \rf{sec4 : prop1} we prove a recursion relation which permits the computation in odd (or even) dimensions of the corresponding kernel in terms of the one- (or two-) dimensional kernels. We also prove precise asymptotics for the kernel at large values of $M$ and $R$ (Euclidean space limit) respectively.  
\par In Sec. \rf{sec5} we study the pointwise convergence at the origin in odd and even dimensions separately because of their distinct behaviour. First we consider the simplest case of functions $f\in C^{\infty}((0,a))$ with zero value elsewhere, and find the necessary and sufficient conditions for convergence. As an application we consider the three dimensional case. Next the more general case of piecewise continuously differentiable functions with compact support is also investigated. The proposed conditions are then applied to an example in five dimensions. To establish convergence in even dimensions we prove the second part of Proposition \rf{sec5 : prop2} using two different methods. The first is based on the Mehler-Fock transformation and the second on the convolution product adjusted to the hyperbolic space. In higher dimensions the recursion relation of Proposition \rf{sec4 : prop1} takes over.            
\section{Preliminaries}
\label{sec1}
\subsection{Geometrical aspects of $\mathbb{H}^d(\mathbb{R})$}
\label{sec1a}
\begin{definition}
The real hyperbolic space $\mathbb{H}^d_{+,R}, \, d\geq 2$ can be realised geometrically as the upper sheet of the two-sheeted hyperboloid embedded into the pseudo-Euclidean space, $\mathbb{E}^{(1,d)}$, namely
\beqr
\mathbb{H}^d_{+,R}=\{X\in \mathbb{E}^{(1,d)} :\,\, [X,X]=R^2, \, X_{0}>0\}\subset \mathbb{R}^{1+d}
\label{sec1a : eq1}
\feqr
where $[\cdot,\cdot]$ is the symmetric bilinear form defined by
\beqr
[X,Y]=X_0Y_0-\sum_{i=1}^{d}X_iY_i=X^tJ_{1,d}Y
\label{sec1a : eq2}
\feqr
with $J_{1,d}$ be the $(1+d)\times(1+d)$ diagonal matrix with signature $(+1,-1,\cdots,-1)$.
\end{definition}
The hyperbolic space $\mathbb{H}^d_{+,R}$ is the maximally symmetric, simply connected, $d$-dimensional Riemannian manifold with constant negative sectional curvature $K=-1/R^2$ for fixed $R$. 
\par We adopt the geodetic spherical polar representation in which the position of a point on the hyperboloid is specified by $(\chi,\theta_j,\phi),\,\chi:=r/R $ with ranges 
\beqr
r\in[0,\infty), \,\, \theta_j\in[0,\pi] \,\,(1\leq j\leq d-2), \,\, \phi\in[0,2\pi).
\label{sec1a : eq5}
\feqr
The geodesic distance between the point $X\in \mathbb{H}^d_{+,R}$ and the pole $\mathcal{O}:=(R,0,\cdots,0)\in \mathbb{H}^d_{+,R}$ is given by \cite{foot1} 
\beqr
r:=R \, d_{\mathbb{H}^d_{+,R}}(X,\mathcal{O})=R\,\cosh^{-1}\left(\frac{[X,\mathcal{O}]}{\sqrt{[X,X]}\sqrt{[\mathcal{O},\mathcal{O}]}}\right)=R\cosh^{-1}\left(\frac{[X,\mathcal{O}]}{R^2}\right).
\label{sec1a : eq4}
\feqr
In Cartesian coordinates a vector $X:=(X^0,\textbf{X})\in \mathbb{R}^+\times\mathbb{R}^{d}$ can be retrieved from $(\chi,\theta_j,\phi)$ by means of the transformations \cite{foot3} 
\beqr
X^{0}&=& R\cosh\chi \non\\
X^1 &=& R\sinh\chi\cos \theta_1 \non \\
X^k &=& R\sinh\chi\left(\prod_{j=1}^{k-1}\sin \theta_j\right) \cos \theta_k, \quad 2\leq k\leq d-2 \nonumber \\
&\vdots& \nonumber \\
X^{d-1}&=& R\sinh\chi\left(\prod_{j=1}^{d-2}\sin \theta_j\right) \cos \phi \nonumber \\
X^d&=& R\sinh\chi\left(\prod_{j=1}^{d-2}\sin \theta_j\right) \sin \phi 
\label{sec1a : eq3}
\feqr
Transformations \rf{sec1a : eq3} can be compactly represented by $X=(R\cosh \chi,R\,\boldsymbol{\xi}\sinh \chi)$, $\boldsymbol{\xi}\in \mathbb{S}^{d-1}$. \\
\textit{Remark 1.} \\
In the $R\rightarrow \infty$ limit the sectional curvature vanishes and \rf{sec1a : eq3} reduce to spherical coordinate transformations which describe the position of a point on an open sphere with center located on the positive $X^0$-axis at infinity, and having infinite radius $r_E$. As an example, consider the geodesics of $\mathbb{H}^{d=2}_{+,R}$ passing through the pole $\mathcal{O}$ projected onto the $X^0=\infty$ plane from the origin of the Cartesian coordinate system. The resulting geodesics on $X^0=\infty$ plane are straight lines emanating from the center of the sphere.
\par The induced line element on $\mathbb{H}^d_{+,R}$ is 
\beqr
ds^2_{\mathbb{H}^d_{+,R}}=R^2 d\chi^2+R^2\sinh^2 \chi \left[\sum_{k=1}^{d-2}\left(\prod_{j=1}^{k-1}\sin^2 \theta_j\right)d\theta_k^2+\left(\prod_{j=1}^{d-2}\sin^2 \theta_j\right)d\phi^2\right]
\label{sec1a : eq6}
\feqr
and the Laplace-Beltrami operator is then given by 
\beqr
\mathcal{L}_{\mathbb{H}^d_{+,R}}&=&\frac{1}{\sqrt{|g|}}\partial_i(\sqrt{|g|}g^{ij}\partial_j)=\frac{1}{R^2}\left(\frac{1}{\sinh^{d-1}\chi}\partial_{\chi}\left(\sinh^{d-1}\chi \partial_{\chi}\right) +\frac{1}{\sinh^2 \chi}\Delta_{\mathbb{S}^{d-1}}\right)\non\\
&=& \mathcal{L}_{R,\chi}+\mathcal{L}_{R,\mathbb{S}^{d-1}}
\label{sec1a : eq7}
\feqr
where $\mathcal{L}_{R,\chi}, \, \mathcal{L}_{R,\mathbb{S}^{d-1}}$ denote the radial and angular part of the Laplacian respectively. 
\par The boundary cone of the unit hypeboloid $\mathbb{H}^d_{+,R=1}:=\mathbb{H}^d_+$ is defined by the set of isotropic vectors for the bilinear form
\beqr
C_+^d=\{p \in \mathbb{R}^{1+d}: [p,p]=0, \, p_0>0 \},
\label{sec1a : eq8}
\feqr
and the projective half null-cone is defined by 
\beqr
\mathbb{P}C_+^d=\{\boldsymbol{p}\in C_+^d: a\boldsymbol{p}\equiv\boldsymbol{p}, \, a>0, p_0>0\}\simeq ~\mathbb{S}^{d-1}=\{\boldsymbol{p}\in \mathbb{R}^d: \parallel\boldsymbol{p}\parallel=1\}
\label{sec1a : eq9}
\feqr
which is diffeomorphic to the $\mathbb{S}^{d-1}$ unit sphere. The
unit sphere in $\mathbb{R}^d$ is identified by the subset
\beqr
\mathcal{B}=\{p \in C_+^d: p_0=1 \}
\label{sec1a : eq10}
\feqr
 via the map
\beqr
\boldsymbol{\omega}\in S^{d-1}\rightarrow p(\boldsymbol{\omega})=(1,\boldsymbol{\omega})\in \mathcal{B}.
\label{sec1a : eq11}
\feqr
\subsection{Hyperbolic plane waves}
\label{sec2a}
On the hyperbolic space we consider the map
\beqr
\mathcal{E}_d: \mathbb{H}^{d}_{+}\times(\mathbb{H}^{d}_{+})^* \rightarrow  \mathbb{C},
\label{sec2a : eq1}
\feqr
where $\mathcal{E}_d(x,p)=[x,p(\boldsymbol{\omega})]^{-\rho+i\lambda}$, $x=(\cosh \chi)\boldsymbol{e_0}+(\sinh\chi)\boldsymbol{\xi}, \,\boldsymbol{\xi}\in \mathbb{S}^{d-1}, \,\rho=\frac{d-1}{2}, \, \lambda=|\textbf{p}|R \in \mathbb{R}^+_0$ and $(\mathbb{H}^{d}_{+})^*=\mathbb{R}^+\times \mathbb{P}C_+^d$ is the dual of the hyperbolic space.  
The map has the following properties:
\begin{description}
\item[$(\alpha)$] It is an eigenfunction of the Laplace-Beltrami operator \rf{sec1a : eq7} with eigenvalue $s(s+2\rho)/R^2$ where $s=-\rho+i\lambda$.
\item[$(\beta)$] The hyperbolic plane waves are exponentially bounded  
\beqr
|\mathcal{E}_d(x,p)|\leq e^{M\rho}, \, 0\leq \chi\leq M, \, 0\leq\theta\leq\pi
\label{sec2a : eq2}
\feqr
and converge to the Euclidean plane waves in the $R\rightarrow \infty$ limit
\beqr
\lim_{R\rightarrow \infty}\mathcal{E}_d(x,p)=e^{i|\textbf{p}|\langle \textbf{r}_E,\boldsymbol{\omega} \rangle}.
\label{sec2a : eq3}
\feqr
\textit{Proof.}\\
Boundedness is easily proved. For the calculation of the limit we have  
\beqr
\lim_{R\rightarrow \infty}\mathcal{E}_d(x,p)&=& \lim_{R\rightarrow \infty}\left(\cosh \chi-\langle \boldsymbol{\xi},\boldsymbol{\omega} \rangle\sinh \chi\right)^{-(\rho+i|\textbf{p}|R)}\non \\
&=&\lim_{R\rightarrow \infty}\left(\cosh \chi \right)^{-(\rho+i|\textbf{p}|R)}\lim_{R\rightarrow \infty}\left(1- \langle \boldsymbol{\xi},\boldsymbol{\omega} \rangle\tanh \chi\right)^{-(\rho+i|\textbf{p}|R)}\non \\
&=&\lim_{R\rightarrow \infty}e^{-i|\textbf{p}| \ln\left(1- \langle \boldsymbol{\xi},\boldsymbol{\omega} \rangle\tanh \chi\right)^R}\non\\
&=&e^{i\langle \textbf{r}_E,\textbf{p}\rangle}
\label{sec2a : eq4} 
\feqr 
where in the last equality we used the $L'H\hat{o}pital's$ rule. This property can also be justified by observing that the one-parameter family of the first fundamental form on $\mathbb{H}^d_{+,R}$ converges to the corresponding one on the Euclidean space, by utilizing the limit $\lim_{R\rightarrow \infty}R \sinh(r/R)=|\textbf{r}_E|$. As a consequence  
\beqr
\lim_{R\rightarrow \infty}(\mathcal{L}_{\mathbb{H}^d_{+,R}})=\mathcal{L}_{\mathbb{E}^d}
\label{sec2a : eq5}
\feqr
and their eigenfunctions are asymptotically identical. 
\item[$(\gamma)$] Under the action of a group element $g\in SO(d)\subset SO_o(1,d)$ (the identity component of $O(1,d)$ preserving the billinear form) the hyperbolic plane wave transforms according to
\beqr
\mathcal{R}(g): \mathcal{E}_d(x,p)\rightarrow \mathcal{E}_d(\mathcal{R}^{-1}(g)x,p(\boldsymbol{\omega}))=\mathcal{E}_d(x,\mathcal{R}(g) p(\boldsymbol{\omega}))
\label{sec2a : eq6}
\feqr
where $\mathcal{R}(g)$ is a representation of $g$.
\end{description}
\subsection{Spherical mean, special functions and the Fourier-Helgason transform}
\label{sec3a}
The spherical mean-value of $\mathcal{E}_d(x,p)$ is defined by
\beqr
\Phi_{\lambda}(\chi)=\frac{1}{|\mathbb{S}^{d-1}|}\int_{\mathbb{S}^{d-1}}\mathcal{E}_d(x,p(\boldsymbol{\omega}))\,d\sigma_{d-1}(\boldsymbol{\omega})
\label{sec3a : eq1}
\feqr 
where $d\sigma_{d-1}(\boldsymbol{\omega})$ is the surface measure on the unit sphere $\mathbb{S}^{d-1}$. $\Phi_{\lambda}(\chi)$ is the unique $SO(d)$ invariant eigenfunction of the radial Laplacian (and thus automatically analytic) satisfying the conditions 
\beqr
\mathcal{L}_{R,\chi}\Phi_{\lambda}(\chi)=-\frac{1}{R^2}(\lambda^2+\rho^2)\Phi_{\lambda}(\chi), \, \, \Phi_{\lambda}(0)=1.
\label{sec3a : eq2}
\feqr
Applying the Funk-Hecke integral identity from Ref. \cite{MO} for the constant spherical harmonic $Y_0(\boldsymbol{\omega})=1$, $\Phi_{\lambda}$ is reduced to a one-variable integral   
\beqr
\Phi_{\lambda}(\chi)&=& \frac{1}{B\left(\rho,\frac{1}{2}\right)}\int_0^{\pi}(\cosh \chi-\cos \theta \sinh \chi)^{-(\rho-i\lambda)}\sin^{2\rho-1}\theta \,d\theta, \, B\left(\rho,\frac{1}{2}\right)=\frac{|\mathbb{S}^{d-1}|}{|\mathbb{S}^{d-2}|} \non\\
&=& 2^{\rho-\frac{1}{2}}\Gamma\left(\rho+\frac{1}{2} \right)\sinh^{-(\rho-\frac{1}{2})}\chi \, P^{-(\rho-\frac{1}{2})}_{-\frac{1}{2}+i\lambda}(\cosh \chi).
\label{sec3a : eq3}
\feqr
The second equality is established in Ref. \cite{ER1} p. 156 where $P^{\mu}_{\nu}(y)$ is the associated Legendre function of the first kind of order $\mu$, degree $\nu$ and argument $y\in (1,\infty), y=\cosh\chi$ .
\par A second realization of $\Phi_{\lambda}(\chi)$ is given in terms of the Gauss hypergeometric function
\beqr
\Phi_{\lambda}(\chi)={}_{2}F_1\left(\frac{1}{2}(\rho+i \lambda),\frac{1}{2}(\rho-i \lambda);\rho+\frac{1}{2};-\sinh^2 \chi\right).
\label{sec3a : eq4}
\feqr
This claim is proved by the definition of the spherical mean and introducing the new integration variable $u=\cos \theta$, thus obtaining
\beqr
\Phi_{\lambda}(\chi)&=&\frac{(\cosh \chi)^{-(\rho-i\lambda)}}{B\left(\rho,\frac{1}{2}\right)}\int_{-1}^{1}(1-u\tanh \chi)^{-(\rho-i\lambda)}(1-u^2)^{\rho-1}du \non \\
&=& \frac{(\cosh \chi)^{-(\rho-i\lambda)}}{B\left(\rho,\frac{1}{2}\right)}\int_{0}^{1}\left[(1-u\tanh \chi)^{-(\rho-i\lambda)}+(1+u\tanh \chi)^{-(\rho-i\lambda)}\right](1-u^2)^{\rho-1}du \non \\
&=&(\cosh \chi)^{-(\rho-i\lambda)} {}_{2}F_1\left(\frac{1}{2}(\rho-i \lambda),\frac{1}{2}(\rho+1-i \lambda);\rho+\frac{1}{2};\tanh^2 \chi\right)\non \\
&=& {}_{2}F_1\left(\frac{1}{2}(\rho + i \lambda),\frac{1}{2}(\rho -i \lambda);\rho+\frac{1}{2};-\sinh^2 \chi\right)
\label{sec3a : eq5}
\feqr
where the formulae 
\beqr
&&\frac{1}{2}(1+\sqrt{z})^{-2a}+\frac{1}{2}(1-\sqrt{z})^{-2a}={}_{2}F_1(a,a+\frac{1}{2};\frac{1}{2};z), \non \\
&&{}_{2}F_1(a,b;c;z)=(1-z)^{-a} {}_{2}F_1(a,c-b;c;\frac{z}{z-1})=(1-z)^{-b} {}_{2}F_1(c-a,b;c;\frac{z}{z-1})\non \\
&& \int_0^1 (1-u^2)^{\rho-1}u^{2n}du=\frac{\Gamma\left(n+\frac{1}{2}\right)\Gamma(\rho)}{2\Gamma\left(n+\rho+\frac{1}{2}\right)}, \,\, n\in \mathbb{Z}^+
\label{sec3a : eq6}
\feqr
from Ref. \cite{ER1} p. 101, 109 and Ref. \cite{RG} p. 343, have been used. Alternatively, this result can also be reproduced by starting from the eigenvalue equation \rf{sec3a : eq2} and performing the transformation $z=-\sinh^2 \chi$. Then the following hypergeometric differential equation, see Ref. \cite{KOO1}, 
\beqr
&& \left(z(1-z)\frac{d^2}{dz^2}+[(a+1)-z(a+b+2)]\frac{d}{dz}\right)\Phi^{(a,b)}_{\lambda}(z)=\frac{1}{4}(\lambda^2+(a+b+1)^2)\Phi^{(a,b)}_{\lambda}(z), \non \\
&& a=\rho-1/2, \,b=-1/2
\label{sec3a : eq7}
\feqr
predicts that the unique regular solution at $z=0$ which satisfies $ \Phi^{(a,b)}_{\lambda}(0)=1$, is given by
\beqr
\Phi_{\lambda}(\chi)&=&\Phi^{(a,b)}_{\lambda}(\chi)={}_{2}F_1\left(\frac{1}{2}(a+b+1+i\lambda),\frac{1}{2}(a+b+1-i\lambda);a+1;-\sinh^2 \chi\right).
\label{sec3a : eq8}
\feqr
This hypergeometric function denotes the unique analytic continuation for $z\not\in [1,\infty)$ of the usual power series expansion of ${}_{2}F_1$ for $|z|<1$. 
\par Next, for completeness, we state without proof, see Refs. \cite{SH1,FJ1,KOO2}, the spherical Fourier transform on the real hyperbolic space. This  theorem will be used in section three.  
\begin{theorem}
\label{sec3a : theo1}
Let $f\in C^{\infty}_0(\mathbb{R}^+)$, $a=\rho-\frac{1}{2}, \, b=-\frac{1}{2}, \, \lambda\in \mathbb{R}$, then the Fourier-Helgason transform is defined by
\beqr
\mathcal{FH}_{a,b}[f](\lambda)=\hat{f}(\lambda):= R^d\int_{\mathbb{R}^+}f(\chi)\Phi_{\lambda}^{(a,b)}(\chi)\sinh^{d-1} \chi d\chi
\label{sec3a : eq9}
\feqr 
and the inverse $\mathcal{FH}^{-1}_{a,b}$ transform is given by 
\beqr
f(\chi)=\frac{2^{2\rho}}{2\pi R^d}\int_{\mathbb{R}^+}\hat{f}(\lambda)\Phi_{\lambda}^{(a,b)}(\chi) |c(\lambda,\rho)|^{-2}d\lambda
\label{sec3a : eq10}
\feqr
where the Harish-Chandra $c$-function is 
\beqr
c(\lambda,\rho)=\frac{2^{\rho-i\lambda}\Gamma(i\lambda)\Gamma\left(\rho+\frac{1}{2}\right)}{\Gamma\left(\frac{1}{2}(\rho+i\lambda)\right)\Gamma\left(\frac{1}{2}(\rho+i\lambda)+\frac{1}{2}\right)}=\frac{2^{2\rho-1}\Gamma(i\lambda)\Gamma\left(\rho+\frac{1}{2}\right)}{\sqrt{\pi}\Gamma(\rho+i\lambda)}.
\label{sec3a : eq11}
\feqr
The mapping \rf{sec3a : eq9} extends to an isomorphism of 
\beqr
L^2([0,\infty), \, \sinh^{d-1}\chi d\chi)\quad onto \quad L^2([0,\infty),\frac{2^{2\rho}}{2\pi R^d}|c(\lambda,\rho)|^{-2}d\lambda ).
\label{sec3a : eq12}
\feqr
\end{theorem}
The second equality in \rf{sec3a : eq12} can be deduced using the Legendre's doubling formula for gamma functions.
\section{The Dirichlet kernel}
\label{sec4}
The d-dimensional Dirichlet kernel on the real hyperbolic space $\mathbb{H}^d_{+,R}$ is defined by the inverse Fourier-Helgason transform of $\textbf{1}_{[0,M]}$, the indicator function of the interval $[0,M]$, namely
\beqr
\mathcal{D}^{(d)}_M(\chi)=\frac{2^{2\rho}}{2\pi R^{d}}\int_{\mathbb{R}^+}\textbf{1}_{[0,M]}(\lambda) \Phi_{\lambda}^{(a,b)}(\chi) |c(\lambda,\rho)|^{-2}d\lambda, \, M=R \tilde{M}
\label{sec4 : eq1}
\feqr
where the factor $R^d$ is required to ensure convergence of the radial hyperbolic measure to the corresponding Euclidean measure. This actually can be  justified by computing the limit (see Appendix for the proof)
\beqr
\lim_{R\rightarrow \infty}(R^{-d}|c(R|\boldsymbol{p}|,\rho)|^{-2} d\lambda)=\frac{4\pi }{2^{4\rho}\Gamma^2\left(\rho+\frac{1}{2}\right)}|\boldsymbol{p}|^{d-1} d|\boldsymbol{p}|.
\label{sec4 : eq2}
\feqr
In one dimension from \rf{sec4 : eq1} we have the elementary trigonometric function
\beqr
\mathcal{D}^{(1)}_M(\chi)=\frac{2\sin (M\chi)}{\pi R \chi}, \quad \chi\neq 0
\label{sec4 : eq3}
\feqr
where $\Phi_{\lambda}^{(-1/2,-1/2)}(\chi)=\cos (\lambda \chi)$ and $|c(\lambda,0)|^2=1/4$ have been used. This is the well-known Shannon's delta kernel or Dirichlet's continuous delta kernel. 
\begin{proposition}
\label{sec4 : prop1} 
The Dirichlet kernel has the following properties:
\begin{description}
\item[$(\alpha)$] It satisfies the recursion relation: 
\beqr D_M^{(d)}(\chi)=-\frac{1}{2a_d R^2\sinh \chi}\frac{d}{d\chi}D_M^{(d-2)}(\chi), \, \, \chi\neq 0, \, \, d=3,4,\cdots, \, a_d=\frac{d-2}{2}.
\label{sec4 : eq4} 
\feqr
\item[$(\beta)$] It converges to the Euclidean Dirichlet kernel 
\beqr
\lim_{R\rightarrow \infty}D_M^{(d)}(\chi)=|\mathbb{S}^{d-1}| \tilde{D}_{\tilde{M}}^{(d)}(|\boldsymbol{r_E}|)
\label{sec4 : eq4a}
\feqr
where the Dirichlet kernel on $\mathbb{R}^d$ is defined by 
\beqr
\tilde{D}_{\tilde{M}}^{(d)}(|\boldsymbol{r_E}|)=\frac{|\mathbb{S}^{d-1}|}{(2\pi)^d}\int_{\mathbb{R}^+} \textbf{1}_{[0,\tilde{M}]}(\chi) \mathcal{J}_a(|\boldsymbol{p}||\boldsymbol{r_E}|)|\boldsymbol{p}|^{d-1}d|\boldsymbol{p}|
\label{sec4 : eq5}
\feqr
with $\mathcal{J}_a$ denoting the spherical Bessel function.
\item[$(\gamma)$] Its asymptotic behaviour for large $M$ is given by
\beqr
D_M^{(d)}(\chi)\sim \frac{2^{-\frac{1}{2}(\rho-3)}}{\sqrt{\pi}\Gamma(\rho+\frac{1}{2})}\frac{\tilde{M}^{\rho}}{\sinh^{\rho+1}\chi}\left[\sin(\tilde{M}r-\frac{\pi \rho}{2})+O(\frac{1}{\tilde{M}})\right].
\label{sec4 : eq6}
\feqr
\item[$(\delta)$] It is an even function, $D_M^{(d)}(\chi)=D_M^{(d)}(-\chi)$.
\item[$(\epsilon)$] At the origin in odd dimensions it has the value  
\beqr
D_M^{(d)}(0)&=&\lim_{\chi\rightarrow 0^+}D_M^{(d)}(\chi)=\frac{1}{2^{2\rho-1}\Gamma^2(\rho+\frac{1}{2})R^d}\int_0^M \frac{|\Gamma(\rho+i\lambda)|^2}{|\Gamma(i\lambda)|^2}d\lambda \non \\
&=&\frac{\Gamma^2(\rho)}{2^{2\rho-1}\Gamma^2(\rho+\frac{1}{2})R^d} \sum_{l=1}^{k}\beta_l\frac{M^{2l+1}}{2l+1}, \, d=2k+1.
\label{sec4 : eq7}
\feqr
\end{description}
\end{proposition}
\textit{Proof.} 
Property $(\alpha)$ follows by applying the identity (see Appendix for the proof)
\beqr
\frac{d}{dz}\Phi_{\lambda}^{(a-1,b)}(z)=\left[\frac{(a+b)^2+\lambda^2}{4a} \right](1-z)^{-(b+1)}{}_{2}F_1\left(\frac{1}{2}(a-b+i\lambda),\frac{1}{2}(a-b-i\lambda);a+1;z\right)
\label{sec4 : eq8}
\feqr
to the Dirichlet kernel, thus having
\beqr
\frac{d}{d\chi}D_M^{(d-2)}(\chi)&=&\frac{2^{2\rho+1}}{\pi}\frac{a^2}{[(a+b)^2+\lambda^2]}\int_0^M \frac{d}{d\chi}\Phi_{\lambda}^{(a-1,b)}(\chi)|c(\lambda,\rho)|^{-2}d\lambda \non \\
&=&-2a R^2 \sinh \chi D_M^{(d)}(\chi).
\label{sec4 : eq9}
\feqr
\par For property $(\beta)$ we first prove the following Lemma.
\begin{lemma}
\label{sec4 : lem1}
The scaled Legendre spherical function, in the $R\rightarrow \infty$ limit, converges to the spherical Bessel function, namely
\beqr
\lim_{R\rightarrow \infty} \mathcal{P}_{-\frac{1}{2}+i\lambda}^{-(\rho-\frac{1}{2})}(\cosh \chi)=\mathcal{J}_{(\rho-\frac{1}{2})}(|\boldsymbol{p}||\boldsymbol{r}_E|), \quad \mathcal{P}_{-\frac{1}{2}+i\lambda}^{-(\rho-\frac{1}{2})}(\cosh \chi)=\sinh^{-(\rho-\frac{1}{2})}\chi P_{-\frac{1}{2}+i\lambda}^{-(\rho-\frac{1}{2})}(\cosh \chi).
\label{sec4 : eq10}
\feqr
\end{lemma}
\textit{Proof.}\\
The spherical Bessel function, $\mathcal{J}_{(\rho-\frac{1}{2})}$, is defined by
\beqr
\mathcal{J}_{(\rho-\frac{1}{2})}(|\boldsymbol{p}||\boldsymbol{r}_E|)=\frac{\int_0^{\pi}e^{i|\boldsymbol{p}||\boldsymbol{r}_E|\cos\theta}\sin^{2\rho-1} \theta\, d\theta}{\int_0^{\pi}\sin^{2\rho-1} \theta\, d\theta}=\frac{2^{\rho-\frac{1}{2}}\Gamma(\rho+\frac{1}{2})}{(|\boldsymbol{p}||\boldsymbol{r}_E|)^{\rho-\frac{1}{2}}}J_{(\rho-\frac{1}{2})}(|\boldsymbol{p}||\boldsymbol{r}_E|)
\label{sec4 : eq11}
\feqr
and in the latter equality, the relation $(8.411.7)$ from Ref. \cite{RG} has been used. 
It follows from $(14.15.13)$ of Ref. \cite{OLV} that for large $\lambda$ and fixed $\rho, \, \chi$, the asymptotic approximation of the Legendre function is obtained by
\beqr
P_{-\frac{1}{2}+i\lambda}^{-(\rho-\frac{1}{2})}(\cosh \chi)&=&\frac{1}{(-\frac{1}{2}+i\lambda)^{\rho-\frac{1}{2}}}\left(\frac{\chi}{\sinh \chi} \right)^{\frac{1}{2}}I_{\rho-\frac{1}{2}}(i\lambda \chi)\left(1+O(\frac{1}{-\frac{1}{2}+i\lambda}) \right), \, \non \\
I_{\rho-\frac{1}{2}}(i\lambda \chi)&=& i^{\rho-\frac{1}{2}}J_{\rho-\frac{1}{2}}(|\boldsymbol{p}||\boldsymbol{|r|})
\label{sec4 : eq12}
\feqr 
uniformly for $\chi\in(0,\infty)$. By noting that $\lim_{\chi\rightarrow 0}(\chi/\sinh \chi)=1$ we recover the desired result. \\
Applying the dominated convergence theorem to \rf{sec4 : eq1}, utilizing the Lemma \rf{sec4 : lem1} and the limit \rf{sec4 : eq2} we end up with the desired result.\\
\textit{Remark 2.} \\
Observe that eq.~\rf{sec3a : eq2}, by making the substitutions $(\chi,\lambda)\rightarrow (r/R,R|\boldsymbol{p}|)$ and letting $R\rightarrow \infty$, is reduced to the Bessel equation
\beqr
u''(|\boldsymbol{r_E}|)+\frac{(2a+1)}{|\boldsymbol{r_E}|}u'(|\boldsymbol{r_E}|)+\tilde{\lambda}^2 u(|\boldsymbol{r_E}|)=0.
\label{sec4 : eq13}
\feqr
In this equation $u(|\boldsymbol{r_E}|)=\mathcal{J}_a(|\boldsymbol{p}||\boldsymbol{r_E}|)=\lim_{R\rightarrow \infty} \Phi_{R |\boldsymbol{p}|}^{(a,-1/2)}(r/R)$ which is the unique even $C^{\infty}$-function obeying $u(0)=1$. Therefore the appearance of Bessel functions is not a surprise.\\
\textit{Remark 3.} \\
An alternative approach, although not general, to prove property $(\beta)$ is by using the series representation of the hypergeometric function ${}_{1}F_2$ 
\beqr
\lim_{R\rightarrow \infty}\Phi_{\lambda}(\chi)&\!\!\!\!=\!\!\!\!&\frac{1}{B\left(\rho,\frac{1}{2}\right)}\int_{0}^{\pi}e^{i|\boldsymbol{r}_E| |\boldsymbol{p}|\cos \theta}\sin^{2\rho-1}\theta d\theta={}_{1}F_2\left(\frac{1}{2};\frac{1}{2},\rho+\frac{1}{2};-\frac{(|\boldsymbol{r}_E| |\boldsymbol{p}|)^2}{4}\right), \, |\boldsymbol{p}||\boldsymbol{r_E}|<2\non\\\
\!\!\!\!&=\!\!\!\!&\Gamma(a+1)\sum_{n=0}^{\infty}\frac{(-1)^n}{n!\Gamma(a+n+1)}\left(\frac{1}{2}|\boldsymbol{p}||\boldsymbol{r_E}|\right)^{2n}\non \\
\!\!\!\!&=\!\!\!\!& \mathcal{J}_a((|\boldsymbol{p}||\boldsymbol{r_E})).
\label{sec4 : eq14}
\feqr
\par Property $(\gamma)$ is based on the asymptotic estimates of $\Phi_{\lambda}(\chi)$ for large $M$ and fixed $R$. In terms of the scaled hypergeometric function ${}_{2}\mathbb{F}_1$, $\Phi_{\lambda}(\chi)$ behaves as in Ref. \cite{JON}
\beqr
\Phi_{\lambda}(\chi)&=&\Gamma(\rho+\frac{1}{2}){}_{2}\mathbb{F}_1\left(\frac{1}{2}(\rho + i \lambda),\frac{1}{2}(\rho -i \lambda);\rho+\frac{1}{2};\frac{1-\cosh(2\chi)}{2} \right)\non \\
&\underset{\lambda \to \infty}{\sim}& \frac{2^{3\rho/2}\sqrt{\chi}}{\sinh^{\rho}\chi}\Gamma(\rho+\frac{1}{2})(i\lambda)^{-(\rho-\frac{1}{2})}I_{\rho-\frac{1}{2}}(i\lambda \chi), \, R-fixed 
\label{sec4 : eq16}
\feqr
and $I_{\nu}(z)$ in its turn behaves as
\beqr
I_{\rho-\frac{1}{2}}(i\lambda \chi)=i^{\rho-\frac{1}{2}}J_{\rho-\frac{1}{2}}(\lambda \chi)
\underset{\lambda \chi \to \infty}{\sim} i^{\rho-\frac{1}{2}}\sqrt{\frac{2}{\pi\lambda \chi}}\left[\cos\left(\lambda \chi-\frac{\pi \rho}{2}\right)+O\left(\frac{1}{\lambda \chi}\right)\right].
\label{sec4 : eq17}
\feqr
Therefore
\beqr
\Phi_{\lambda}(\chi)\sim \frac{2^{(3\rho+1)/2}\Gamma(\rho+\frac{1}{2})}{(\lambda\sinh \chi)^{\rho}\sqrt{\pi}}\left[\cos\left(\lambda \chi-\frac{\pi \rho}{2}\right)+O\left(\frac{1}{\lambda \chi}\right)\right].
\label{sec4 : eq18}
\feqr
Also $|c(\lambda,\rho)|^{-2}$ behaves as
\beqr
|c(\lambda,\rho)|^{-2}\underset{\lambda \to \infty}{\sim} \frac{\pi}{2^{2(2\rho-1)}\Gamma^2(\rho+\frac{1}{2})}\lambda^{2\rho}.
\label{sec4 : eq19}
\feqr
Combining the definition of the Dirichlet kernel \rf{sec4 : eq1} and the asymptotics \rf{sec4 : eq18} and \rf{sec4 : eq19} we complete the proof. \\
\textit{Remark 4.} \\
This property can be used to prove that the Dirichlet kernel is not a good kernel since the integral of its absolute value is large. 
\par Property $(\delta)$ is immediate from the definition of $D_M^{(d)}(\chi)$. \par The last property which holds only in odd dimensions can be verified using the Appendix. We have not recorded the result in the even-dimensional case since the corresponding expression is lengthy and depends on the value of $d$.
\section{Convergence at the origin}
\label{sec5}
We study first the $d=2k+1$ case. It is known that the sequence $\{\delta_M(\chi)=\frac{\sin (M\chi)}{\pi \chi}\}$ of distributions converges in the weak-$\star$ topology to the Dirac function in the $M\rightarrow \infty$ limit. Applying the recurrent identity \rf{sec4 : eq4} we find that 
\beqr
D^{(2k+1)}_M(\chi)=\frac{(-1)^k}{2^{k-1}R^{2k+1} \prod_{m=0}^{k-1}a_{d-2m}}\mathcal{\hat{A}}^k \delta_M(\chi), k\in\mathbb{N}
\label{sec5 : eq1}
\feqr
where $\mathcal{\hat{A}}=\frac{\partial}{\partial(\cosh \chi)}$. 
\begin{proposition}
\label{sec5 : prop1}
Let $f\in C^{\infty}((0,a))$, $\lim_{\chi\rightarrow 0+}f^{(m)}(\chi)=0; \, m=0,\cdots,k$ and $\lim_{\chi\rightarrow a^-}f^{(l)}(\chi)=0; \, l=0,\cdots,k-1$ where $f^{(m)}(\chi)=\frac{d^m f(\chi)}{d\chi^m}$. Then the spherical partial sum of the Fourier transform converges to $f(0^+)$, namely, 
\beqr
f(0^+)=R^d \lim_{M\rightarrow \infty}\int_0^a f(\chi)D^{(d)}_M(\chi)\sinh^{d-1}\chi d\chi.
\label{sec5 : eq2}
\feqr
\end{proposition}
\textit{Proof.} \\
Substituting \rf{sec5 : eq1} into \rf{sec5 : eq2} and integrating by parts we obtain  
\beqr
B_k\lim_{M\rightarrow \infty}\!\!\!\!&\Biggl(&\!\!\!\!\sum_{l=1}^{k}(-1)^{l-1}(\mathcal{\hat{A}}^{k-l}\delta_M(\chi))\mathcal{\hat{A}}^{l-1}(f(\chi)\sinh^{2k-1}(\chi)) \Biggr|_{\chi=0}^{a} \label{sec5 : eq3} \\
&+&(-1)^k \int_0^a \delta_M(\chi)\partial_{\chi}(\mathcal{\hat{A}}^{k-1}(f(\chi)\sinh^{2k-1}(\chi)))d\chi\Biggr)=f(0^+), \non\\
B_k&=&\frac{(-1)^k}{2^{k-1}\prod_{m=0}^{k-1}a_{d-2m}}=\frac{2(-1)^k}{(2k-1)!!}.
\non
\feqr
In the previous calculation we used the following integral identity and limits
\beqr
\int_0^{\infty}\frac{\sin u}{u}du=\frac{\pi}{2}, \, \lim_{\chi\rightarrow 0^+}\delta_M^{(2l)}(\chi)=\frac{(-1)^k M^{2l+1}}{(2l+1)\pi},\, \lim_{\chi\rightarrow 0^+}\delta_M^{(2l+1)}(\chi)=0.
\label{sec5 : eq4}
\feqr
The elimination of the boundary terms in \rf{sec5 : eq3} can be achieved by imposing the conditions of the proposition \rf{sec5 : prop1}. Taking into account that only the term $(2^k/\pi)\prod_{m=0}^{k-1}a_{d-2m}\cosh^k \chi f(\chi)$ from $\partial_{\chi}(\mathcal{\hat{A}}^{k-1}(f(\chi)\sinh^{2k-1}(\chi)))$ contributes, we reach the final result. \\
\textit{Example 1.} \\
The Dirichlet kernel in $d=3$ is found to be
\beqr
D^{(3)}_M(\chi)=-\frac{2}{R^3\sinh \chi}\mathcal{\hat{A}}\delta_M(\chi)=-\frac{2}{\pi R^3\sinh \chi}\left(\frac{M\chi \cos(M\chi)-\sin(M\chi)}{\chi^2}\right).
\label{sec5 : eq5}
\feqr 
Substituting it into \rf{sec5 : eq2} and partial integrating we have
\beqr
R^3 \lim_{M\rightarrow \infty}\int_0^a f(\chi)D^{(3)}_M(\chi)\sinh^{2}\chi d\chi&=&\frac{2}{\pi}\lim_{M\rightarrow \infty}\left(\int_0^a \frac{d}{d\chi}(f(\chi)\sinh \chi)\delta_M(\chi)d\chi\right) \non \\
&=& \frac{2}{\pi}f(0^+)\int_0^{\infty}\frac{\sin u}{u}du=f(0^+)
\label{sec5 : eq6}
\feqr
where we used $f(a^-)=0$ and the dominated convergence theorem. \\
\textit{Remark 5.} \\
Proposition \rf{sec5 : prop1} can be extended to piecewise continuously differentiable functions $f$ defined on the interval $(0,a)$ with the property, f is of class $C^{k-1}$ on each subinterval $(b_{i-1},b_i), \, i=1,\cdots,n$ of the finite partition $0=b_0<b_1<\cdots<b_n=a$. Also the one-sided limits $\lim_{\chi\rightarrow b_{i}^-}f^{(l)}(\chi)$, $\lim_{\chi\rightarrow b_{i}^+}f^{(l)}(\chi)$ exist for every $i=1,\cdots,n-1, \, l=1,\cdots,k-1$ and $\lim_{\chi\rightarrow b_{0}^+}f^{(l)}(\chi), \,\lim_{\chi\rightarrow b_{n}^-}f^{(l)}(\chi)$ also exist. \\
\textit{Example 2.} \\
Motivated by the previous remark we study the $d=5$ case. Consider the finite partition $0=b_0<b_1<\cdots<b_n=a$ and a function $f$ fulfilling the above conditions. Then the Dirichlet kernel is given by 
\beqr
D^{(5)}_M(\chi)=\frac{1}{R^5 a_3 a_5}\mathcal{A}^2\delta_M(\chi)=-\frac{2}{3R^5}\left(\frac{\cosh \chi}{\sinh^3 \chi}\frac{\partial \delta_M(\chi)}{\partial \chi}-\frac{1}{\sinh^2 \chi}\frac{\partial^2 \delta_M(\chi)}{\partial \chi^2}\right)
\label{sec5 : eq7}
\feqr
and substituting it into the right-hand side of \rf{sec5 : eq2} and partial integrating we have
\beqr
&&R^5\!\!\! \lim_{M\rightarrow \infty}\int_0^a \!\!\!\!f(\chi)D^{(5)}_M(\chi)\sinh^{4}\!\!\chi d\chi\!\!=\!\!\frac{2}{3}\lim_{M\rightarrow \infty}\Biggl[\frac{1}{2}\sum_{i=1}^{n-1}\!\delta_M(b_i)\delta f(b_i)\sinh(2b_i)-\frac{1}{2}\delta_M(a^-)f(a^-)\sinh(2a^-)\non\\
\!\!\!\!&-&\!\!\!\!\sum_{i=1}^{n-1}\!\frac{\partial \delta_M(b_i)}{\partial \chi}\delta f(b_i)\sinh^2(b_i)+\frac{\partial \delta_M(a^-)}{\partial \chi}f(a^-)\sinh^2(a^-)-\sum_{i=1}^{n-1}\!\delta_M(b_i)(\delta f(b_i)\sinh(2b_i)+\delta f^{(1)}(b_i)\sinh^2(b_i))\non \\
\!\!\!\!&+&\!\!\!\!\delta_M(a^-)(f(a^-)\sinh(2a^-)+f^{(1)}(a^-)\sinh^2(a^-))+\frac{1}{2}\int_0^a \delta_M(\chi)\frac{d}{d\chi}(f(\chi)\sinh(2\chi))d\chi\non \\
\!\!\!\!&+&\!\!\!\!\int_0^a \delta_M(\chi)\frac{d^2}{d\chi^2}(f(\chi)\sinh^2 \chi)d\chi\Biggr]
\label{sec5 : eq8}
\feqr
where $\delta f^{(m)}(b_i)=f^{(m)}(b_i^+)-f^{(m)}(b_i^-)$. Piecewise smoothness implies the vanishing of the boundary terms which are proportional to $\delta f(b_i)$ and the last two integrals both contribute to produce the final result.   
\par Next we investigate the even-dimensional case which will be treated similarly. In $d=2$ we have
\beqr
\Phi_{\lambda}^{(0,-\frac{1}{2})}(\chi)=P_{-\frac{1}{2}+i\lambda}(\cosh \chi),\, |c(\lambda)|^{-2}=\pi \lambda \tanh(\pi \lambda)
\label{sec5 : eq9}
\feqr 
where $P_{-\frac{1}{2}+i\lambda}(\cosh \chi)$ is the conical or Mehler's function of zero order, see Ref. \cite{ZH}. The Dirichlet kernel is then written as
\beqr
D_M^{(2)}(y)=\frac{1}{R^2}\int_0^M P_{-\frac{1}{2}+i\lambda}(y)\lambda \tanh(\pi \lambda)d\lambda=\frac{1}{R^2}\delta_M(y), \, y=\cosh \chi, \, 1\leq y<\infty.
\label{sec5 : eq10}
\feqr
\begin{proposition}
\label{sec5 : prop2}
Let $f(y)$ be a continuous function of bounded variation in any finite interval of $(1,\infty)$ and such that 
\beqr
\int_1^{\infty} \frac{f(y)}{\sqrt{1+y}}dy<\infty
\label{sec5 : eq12}
\feqr
then it holds 
\beqr
\mathfrak{M F}[g](y)=f(y)=\int_{0}^{\infty}P_{-\frac{1}{2}+i\mu}(y)g(\mu)d\mu
\label{sec5 : eq13}
\feqr
where $\mathfrak{M F}[g](y)$ is the Mehler-Fock transformation of $g$ with inverse 
\beqr
g(\mu)=\mu \tanh(\pi \mu)\int_1^{\infty}P_{-\frac{1}{2}+i\mu}(y)f(y)dy.
\label{sec5 : eq14}
\feqr
Moreover, 
\beqr
\lim_{M\rightarrow \infty}\left(\int_1^{\infty}f(y)\delta_M(y)dy\right)=f(1^+).
\label{sec5 : eq15}
\feqr 
\end{proposition}
\textit{Proof.} \\
The first part of proposition is a version of a theorem stated in Ref. \cite{F}. For the second part, substituting the Mehler-Fock transformation of $g$ into the left-hand side of \rf{sec5 : eq15}, using the sequence $\{\delta_{M}(y)\}$ and the dominated convergence theorem, we have after integral rearrangements 
\beqr
\int_0^{\infty}g(\mu)\int_0^{\infty}\lambda \tanh(\pi \lambda)\left(\int_1^{\infty}P_{-\frac{1}{2}+i\mu}(y)P_{-\frac{1}{2}+i\lambda}(y)dy\right)d\lambda d\mu.
\label{sec5 : eq16}
\feqr
The completeness relation in Ref. \cite{NOR}
\beqr
\int_1^{\infty}P_{-\frac{1}{2}+i\mu}^m(y)P_{-\frac{1}{2}+i\lambda}^m(y)dy=\frac{(-1)^m}{\mu \tanh(\pi\mu)}\frac{\Gamma(\frac{1}{2}+m +i\mu)}{\Gamma(\frac{1}{2}-m +i\mu)}\delta(\mu-\lambda)
\label{sec5 : eq17}
\feqr
with $m=0$, reduces the triple integral \rf{sec5 : eq16} to a single one, and thus obtaining from \rf{sec5 : eq13} the relation
\beqr
\int_0^{\infty}g(\mu)d\mu=\lim_{y\rightarrow 1^+}\left(\int_{0}^{\infty}P_{-\frac{1}{2}+i\mu}(y)g(\mu)d\mu \right)=f(1^+)
\label{sec5 : eq18}
\feqr
recalling that $\lim_{y\rightarrow 1^+}P_{-\frac{1}{2}+i\lambda}(y)=1$. 
\par An alternative approach to prove the aforementioned claim is to verify in two dimensions the following limit \cite{foot2}
\beqr
\lim_{M\rightarrow \infty}(f\ast D_M^{(2)})(x)=\frac{1}{\pi R^2}f(x)
\label{sec5 : eq19}
\feqr 
where $\ast$ is the convolution product of two spherical functions defined by 
\beqr 
(f\ast g)(x)=\int_{1}^{\infty}f(y)(T_x g)(y)\, dy.
\label{sec5 : eq20}
\feqr
In \rf{sec5 : eq20} $T_x$ is the generalized translation operator which for spherical functions and $d=2$ dimensions, see Ref. \cite{KOO3}, is defined by 
\beqr 
(T_x g)(y)&=&\frac{1}{\pi}\int_0^{\pi}g(xy+\sqrt{(x^2-1)(y^2-1)}\cos \theta)\, d\theta=\int_{z_1}^{z_2}K(x,y,z)g(z)dz, \non \\
z&=&xy+\sqrt{(x^2-1)(y^2-1)}\cos \theta)
\label{sec5 : eq21}
\feqr 
where $z_1=xy-\sqrt{(x^2-1)(y^2-1)}$, $z_2=xy-\sqrt{(x^2-1)(y^2-1)}$ and
\beqr
K(x,y,z)=\left\{\begin{array}{cc} \frac{1}{\pi}\frac{1}{\sqrt{(z-z_1)(z_2-z)}},& z\in(z_1,z_2)\\
0, & otherwise.
\end{array} \right.
\label{sec5 : eq22}
\feqr
Starting from the right hand side of \rf{sec5 : eq19} and using Theorem \rf{sec3a : theo1} in two dimensions we have
\beqr
f(x)&=&\frac{1}{\pi}\lim_{M\rightarrow \infty}\int_1^{\infty}f(y)dy \left( \int_0^{M}P_{-\frac{1}{2}+i\mu}(x)P_{-\frac{1}{2}+i\mu}(y)\pi \lambda \tanh(\pi \lambda)\, d\lambda \right)\non \\
&=&\frac{1}{\pi}\lim_{M\rightarrow \infty}\int_1^{\infty}f(y)dy \left( \int_0^{M}\left(\int_1^{\infty}K(x,y,z)P_{-\frac{1}{2}+i\mu}(z)dz\right)\pi \lambda \tanh(\pi \lambda)\, d\lambda \right) \non \\
&=&\pi R^2 \lim_{M\rightarrow \infty}\int_1^{\infty}f(y)(T_x D_M^{(2)})(y)dy\non \\
&=& \pi R^2 \lim_{M\rightarrow \infty}(f\ast D_M^{(2)})(x).
\label{sec5 : eq23}
\feqr  
In the second equality we used the product formula from Ref. \cite{VIL}
\beqr
P_{-\frac{1}{2}+i\mu}(x)P_{-\frac{1}{2}+i\mu}(y)=\int_1^{\infty}K(x,y,z)P_{-\frac{1}{2}+i\mu}(z)dz.
\label{sec5 : eq24}
\feqr
For $d=2k$ and $k\geq 2$ proposition \rf{sec5 : prop1} still holds. 
\section*{Conclusions}
In this paper we proposed a unified framework to study the asymptotic behaviour and the pointwise convergence of the spherical Dirichlet kernel, at a pre-assigned point,     on the real hyperbolic space. The basic idea of this method relies on the position of the pole of the hyperboloid, in geodetic spherical polar coordinates, which at infinity enables the transition from hyperbolic to Euclidean geometry.  
\section*{Acknowledgements}
The author would like to thank  A. B. Olde Daalhuis (Edinburgh University, School of Mathematics, United Kingdom) for valuable comments on asymptotic expansions.
\section*{Appendix A}
\label{apC}
\renewcommand{\theequation}{A.\arabic{equation}}
\setcounter{equation}{0}
\textbf{Proof of \rf{sec4 : eq2}}\\
The gamma function satisfies the following identities (8.332.1) and (8.332.2) of Ref. \cite{RG}
\beqr |\Gamma(i\lambda)|^2&=&\frac{\pi}{\lambda \sinh (\pi \lambda)},\quad \lambda\in\mathbb{R} \non \\
\left|\Gamma\left(\frac{1}{2}+i\lambda\right)\right|^2&=& \frac{\pi}{\cosh (\pi \lambda)}\non \\
|\Gamma(k+i\lambda)|^2&=&|\Gamma(i\lambda)|^2\prod_{l=0}^{k-1}(l^2+\lambda^2), \quad k\in \mathbb{N} \non \\
\left|\Gamma\left(k+\frac{1}{2}+i\lambda\right)\right|^2&=&\left|\Gamma\left(\frac{1}{2}+i\lambda\right)\right|^2\, \prod_{l=0}^{k-1}\left[(l+\frac{1}{2})^2+\lambda^2\right].
\label{apC : eq1}
\feqr
The last two identities can be proved directly from the first two. The modulus squared of the Harish-Chandra $c-$function is given by
\beqr
|c(\lambda,\rho)|^2=|c(-\lambda,\rho)|^2=\left\{ \begin{array}{ll}
\frac{4^{2k-1}\Gamma^2\left(k+\frac{1}{2}\right)}{\pi \prod_{l=0}^{k-1}(l^2+\lambda^2)}, & d=2k+1 \\
& \\
\frac{4^{2(k-1)}\Gamma^2(k)}{\pi \lambda \tanh(\pi \lambda)\prod_{l=0}^{k-2}\left[(l+\frac{1}{2})^2+\lambda^2\right]}, & d=2k. 
\end{array} \right.
\label{apC : eq2}
\feqr
In the previous expressions the products equal unity for $k=0$ in odd dimensions and for $k=1$ in even dimensions. The finite products can be rewritten as polynomials in $\lambda$ for $d=2k+1$ and $d=2k$ respectively. Namely,
\beqr
P_{k-1}(\lambda)&\equiv& \prod_{l=0}^{k-1}(l^2+\lambda^2)=\Gamma^2(k)\left(\beta_1 \lambda^2+\beta_2 \lambda^4+\cdots + \beta_k \lambda^{2k}  \right), \quad \textrm{where} \non \\
\beta_1&=& 1, \quad \beta_j=\sum_{n_{j-1}=1;\cdots;n_2=n_3+1;n_1=n_2+1 }^{k-j;\cdots;k-2;k-1}\frac{1}{n_1^2 n_2^2\cdots n_{j-1}^2}, \,\, 2\leq j\leq k-1, \,\, \beta_k=\frac{1}{\Gamma^2(k)}
\label{apC : eq3}
\feqr
and
\beqr
P_{k-2}(\lambda)&\equiv& \prod_{l=0}^{k-2}\left[(l+\frac{1}{2})^2+\lambda^2\right]=\alpha_0\left(1+\beta_1 (2\lambda)^2+\cdots+\beta_{k-1}(2\lambda)^{2(k-1)}\right), \quad \textrm{where}\non \\
\alpha_0&=&\frac{1}{\pi}\Gamma^2\left(k-\frac{1}{2} \right), \quad \beta_{k-1}=\frac{1}{2^{2(k-1)}\alpha_0}\non \\ 
\beta_j&=&\sum_{n_{j}=0;\cdots;n_2=n_3+1;n_1=n_2+1 }^{k-(j+1);\cdots;k-3;k-2}\frac{1}{(2n_1+1)^2 (2n_2+1)^2\cdots (2n_{j}+1)^2}, \quad 2\leq j\leq k-2.
\label{apC : eq4}
\feqr 
Then it is evident that only the highest degree term in $\lambda$ contributes a non-vanishing result in the $R\rightarrow \infty$ limit.\\
\textbf{Proof of \rf{sec4 : eq8}} \\
Applying the identity, see, e.g. Ref. \cite{ER1} p. 102, 
\beqr
\frac{d}{dz}{}_{2}F_1\left(\alpha,\beta;\gamma;z\right)=\frac{(\alpha)_1 (\beta)_1}{(\gamma)_1}{}_{2}F_1\left(\alpha+1,\beta+1;\gamma+1;z\right)
\label{apC : eq5}
\feqr
to $\Phi_{\lambda}^{(a-1,b)}(z)$ we obtain
\beqr
\frac{d}{dz}\Phi_{\lambda}^{(a-1,b)}(z)=\left[\frac{(a+b)^2+\lambda^2}{4a} \right]\Phi_{\lambda}^{(a,b+1)}(z).
\label{apC : eq6}
\feqr
Additionally, using the identity 
\beqr
{}_{2}F_1\left(\alpha,\beta;\gamma;z\right)=(1-z)^{\gamma-\alpha-\beta}{}_{2}F_1\left(\gamma-\alpha,\gamma-\beta;\gamma;z\right)
\label{apC : eq7}
\feqr
from \cite{RG} p. 1069, we finally recover the desired expression.


\begin{thebibliography} {00}
\bibitem{foot1} The geogesic distance between two points on the hypersphere of radius $R$ is given by 
\[d_{\mathbb{S}^d_R}(X,Y)=R\cos^{-1}\left(\frac{[X,Y]}{R^2}\right), \, \forall X,Y\in \mathbb{S}^d_R\]
where $[X,Y]=X_0Y_0+\sum_{i=1}{d}X_iY_i$ is the Euclidean inner product.
\bibitem{foot2} In what follows $x,y$ and $z$ are expressed in terms of the hyperbolic cosine functions of the corresponding geodetic distance. For example $x=\cosh \chi$.
\bibitem{foot3} If the coordinate $X^0$ represents time in $\mathbb{E}^{(1,2)}$ then these transformations can be derived by applying to the pole a composition of a pure rotation with a pure boost. Both are elements of the restricted Lorentz group $\mathcal{L}_+^{\uparrow}$ (the orthochronous, proper Lorentz group). The pure boost $B(\boldsymbol{e_1},\chi)$ in the direction $\boldsymbol{e_1}$, the unit vector in the direction $X^1$, and with rapidity parameter $\chi$ is given by the matrix
\begin{equation*}
B(\boldsymbol{e_1},\chi)=
\begin{pmatrix}
\cosh \chi & \sinh \chi & 0 \\
\sinh \chi & \cosh \chi & 0 \\
0 & 0 & 1
\end{pmatrix}
\end{equation*}   
and the pure rotation $\mathcal{R}(\boldsymbol{e_0}, \theta)$ about the axis $\boldsymbol{e_0}$ by the angle $\theta$ is given by
\begin{equation*}
\mathcal{R}(\boldsymbol{e_0},\theta)=
\begin{pmatrix}
1 & 0 & 0 \\
0 & \cos \theta & -\sin \theta \\
0 & \sin \theta & \cos \theta
\end{pmatrix}.
\end{equation*} 
The point $P\in \mathcal{H}^2_{+,R}$ is then connected to the pole $\mathcal{O}$ through the transformation \[P=\mathcal{R}(\boldsymbol{e_0},\theta)B(\boldsymbol{e_1},\chi)\mathcal{O}.\]
\end{thebibliography}
\end{document}